\documentclass[a4paper,fleqn,longmktitle]{cas-dc}
\usepackage[numbers]{natbib}
\usepackage{gensymb}
\usepackage{titlesec}
\usepackage{color}
\usepackage{verbatim}
\usepackage{hyperref}
\usepackage{setspace}
\titleformat*{\subsection}{\normalsize\itshape}

\titleformat*{\section}{\normalsize\bfseries}

\let\printorcid\relax

\begin{document}

\let\WriteBookmarks\relax
\def\floatpagepagefraction{1}
\def\textpagefraction{.001}
\shorttitle{Property investigation for different wedge-shaped CsI(Tl)s}
\shortauthors{Gen L et~al.}

\title [mode = title]{Property investigation for different wedge-shaped CsI(Tl)s}                      


\author[1]{G. Li}[]

\address[1]{School of Physics and State Key Laboratory of Nuclear Physics and Technology, Peking University, Beijing 100871, China}

\author[1]{J. L. Lou}[]
\cormark[1]
\cortext[cor1]{Corresponding author.
E-mail Address: jllou@pku.edu.cn}
\author[1]{Y. L. Ye}[]%
\author[1]{H. Hua}[]%
\author[1]{H. Wang}
\author[1]{J. X. Han}[]%
\author[1]{W. Liu}[]%
\author[1]{S. W. Bai}[]%
\author[1]{Z. W. Tan}[]%
\author[1]{K. Ma}[]%
\author[1]{J. H. Chen}[]%
\author[1]{L. S. Yang}[]%
\author[1]{S. J. Wang}[]%
\author[1]{Z. Y. Hu}[]%
\author[1]{H. Z. Yu}[]%
\author[1]{H. Y. Zhu}[]%
\author[1]{B. L. Xia}[]%
\author[1]{Y. Jiang}[]%
\author[1]{Y. Liu}[]%
\author[1]{X. F. Yang}[]%
\author[1]{Q. T. Li}[]%
\author[1]{J. Y. Xu}[]%
\author[2,3]{J. S. Wang}[]%
\author[2]{Y. Y. Yang}[]%
\author[2]{J. B. Ma}[]%
\author[2]{R. F. Chen}[]%
\author[2]{P. Ma}[]%
\author[2]{Z. Bai}[]%
\author[2]{F. F. Duan}[]%
\address[2]{Institute of Modern Physics, Chinese Academy of Sciences, Lanzhou, 730000, China}
\address[3]{School of Science, Huzhou University, Huzhou, 313000, China}  
\author[4]{L. Y. Hu}[]%
\author[4]{J. W. Li}[]%
\author[4]{Y, Li}[]%
\author[4]{Y. S. Song}[]%
\address[4]{Fundamental Science on Nuclear Safety and Simulation Technology Laboratory, Harbin Engineering University, Harbin, 1500001, China}
\author[5]{Suyalatu Zhang}[]%
\author[5]{M. R. Huang}[]%

\address[5]{Institute of Nuclear Physics,
Inner Mongolia University for Nationalities, Tongliao, 028000, China}

\begin{abstract}
Two types of wedge-shaped CsI(Tl)s were designed to be placed behind the annular double-sided silicon detectors (ADSSDs) to identify the light charged particles with the $\Delta E-E$ method. The properties of CsI(Tl)s with different shapes and sizes, such as energy resolution, light output non-uniformity and particle identification capability, were compared by using a $\alpha$-source and a radioactive beam of $^{15}$C. The big-size CsI(Tl) was finally adopted to form the $\Delta E-E$ telescope due to better properties. The property differences of these two types of CsI(Tl)s can be interpreted based on the Geant4 simulation results. 
\end{abstract}



\maketitle
\doublespacing

\section{Introduction}
\begin{figure}
	\centering
		\includegraphics[scale=.25]{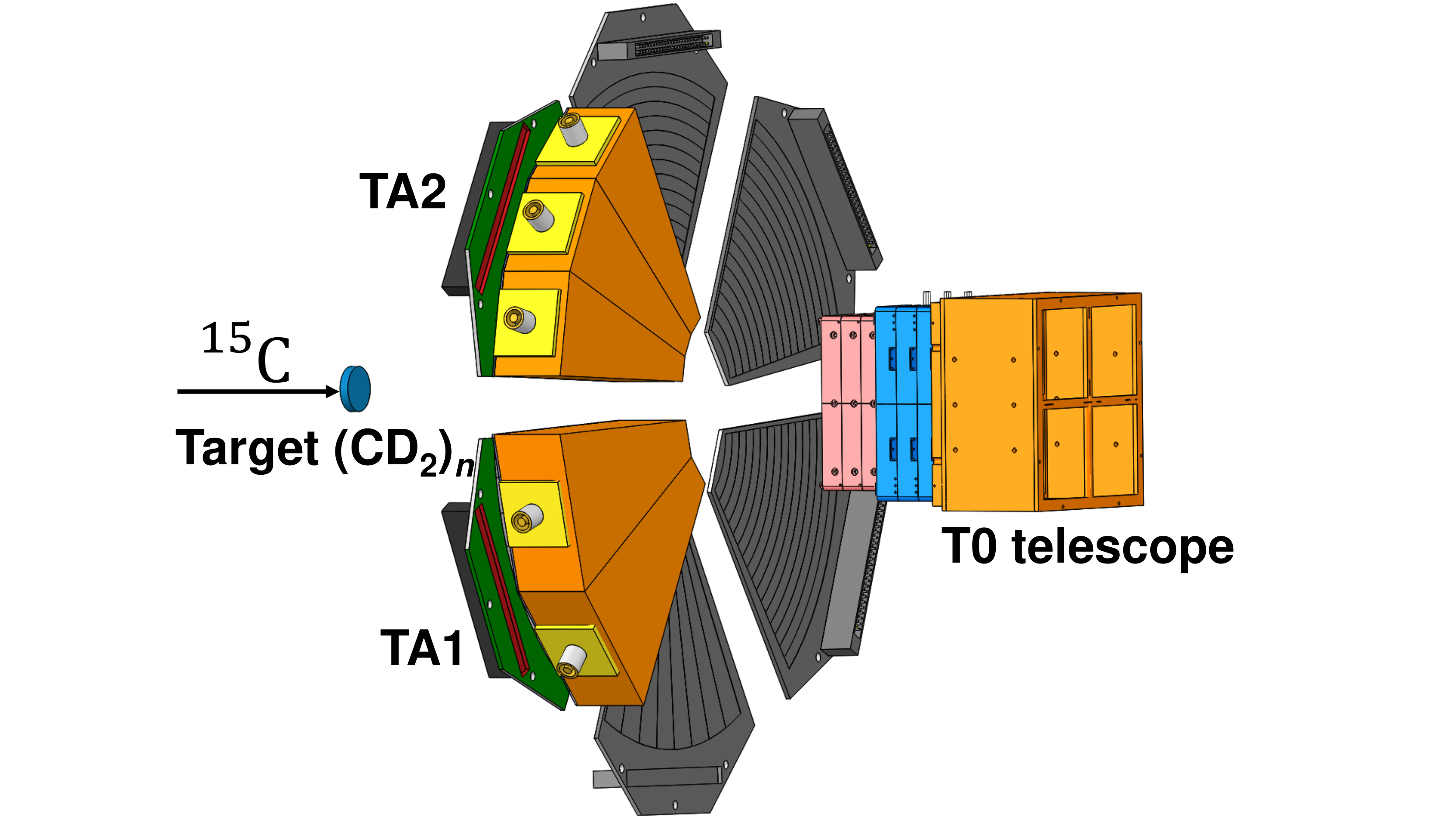}
	\caption{Schematic view of the telescope TA, which is used to detect the light charged particles from reactions induced by a secondary beam impinging on a physical target. The telescope T0 is installed around 0$^\circ$ relative to the secondary beam direction to measure the heavy residuals.}
	\label{setup}
\end{figure}

A silicon detection array was constructed at Peking University in order to study exotic structures of unstable nuclei via direct nuclear reaction experiments~\cite{Liuwei2020,ligen2020}. The array has been successfully used in several experiments to identify the charged particles by using both TOF - $\Delta E$ (time-of-flight and energy deposit) and $\Delta E$ - $E$ methods \cite{Liu2020,jiang2020quadrupole,Yang2014,ChenPRC,Chenplb}. One of the most important parts is a set of annular double-sided silicon strip detectors (ADSSD). As shown in Fig.~\ref{setup}, it has large sensitive area, and thus covers large solid angles with an acceptable angular resolution. According to the kinematics, it was placed upstream of the physical target in the ($d$, $p$) reaction to detect protons~\cite{ChenPRC,Chenplb}, and installed downstream in ($d$, $^3$He), ($d$, $t$), ($d$, $d$’), and ($p$, $p$’) reactions to discriminate $^3$He, $t$, $d$, and $p$, respectively~\cite{Liu2020,jiang2020quadrupole}. It was found that the light mass nuclei with different charge number ($Z$) can be clearly distinguished through the TOF - $\Delta E$ method, where TOF is the time of flight between the plastic scintillation detector installed in the beam line and the ADSSD, and $\Delta E$ is the energy deposited in the ADSSD ~\cite{Liu2020,jiang2020quadrupole}. However, this method is difficult to identify isotopes with the same $Z$ produced from reactions induced by a secondary beam impinging on a physical target due to the limited time resolution of silicon detector and the short time-of-flight in the vacuum chamber. This prompts us to design a new telescope, consisting of the ADSSD (as a $\Delta E$ detector) and another $E$ detector, to distinguish light charged particles via the $\Delta E - E$ method. 

The thallium-activated caesium iodide crystal (CsI(Tl)) scintillation detectors are widely used in nuclear physics experiments. Due to a high density  ($\rho \thickapprox 4.5 g/\rm{cm}^3$), CsI(Tl) with thin thickness can be used to stop charged particles. Thus, it is suitable for collecting residual energy after silicon detectors based on the telescope system to identify particles via the $\Delta E-E$ technology. The photo-diode (PD), which has not only a small size but also good performance in photon collection and photoelectric conversion, is appropriate for the read-out of the CsI(Tl) detector behind the ADSSD. However, the light output non-uniformity of the polished tapered crystals were reported in Refs. \cite{murray1961scintillation,bea1994simulation,knyazev2019properties,AUFFRAY200222}, which decreases the energy resolution of crystals and further affects the particle-identification (PID) capability of the $\Delta E-E$ method.
In this paper, we designed two types of wedge-shaped CsI(Tl)s to match the ADSSD size, then tested and compared their energy resolution, PID capability, and light output non-uniformity by using a $\alpha$-source and a radioactive beam of $^{15}$C. Finally, we chose one shape with better properties as the $E$ detector of the telescope TA.

\section{Details of CsI(Tl)}\label{experimentset}
A set of ADSSD is comprised of six wedge-shaped parts, each has an inner (outer) radius of 32.6 mm (135.1 mm) and a central angle of nearly 56$^\circ$ \cite{Chenplb}. Each part is divided into sixteen 6.4-mm-wide ring-strips on one side and 8 wedge-shaped regions on the other side. Two different shapes which can match the ADSSD size are designed. As shown in Fig.~\ref{setup}, three smaller or two bigger crystals are assembled as a $E$ detector, and then placed behind a ADSSD with a thickness of $\sim$150 $\micro$m. 

 The thickness of each CsI(Tl) is designed to be 31 mm in order to stop the charged particles up to 80 MeV that we desire to detect. The height of all the crystals are needed to be more than 10 cm for the purpose of covering the size of ADSSD. As shown in Fig.~\ref{CsI} (a), each crystal with the same thickness and height looks like a flat-topped pyramid. Four peripheral surfaces are rectangular, while the front and back surfaces are isosceles trapezoid. The length of longer and shorter side in isosceles trapezoid is 41.2 and 7.0 mm for the smaller scintillator, and 66.4 and 13.2 mm for the bigger one, respectively. All the surfaces of each CsI(Tl) are polished. Four of six surfaces of each CsI(Tl) are packed by the tyvek paper except the front surface faced to the ADSSD and the readout surface. The front (particle incident) surface is packed by a 2-$\mu$m aluminized Mylar foil to let charged particles punch through easily and to reflect fluorescent lights back into scintillators. The readout surface is coupled to a PD with silicone grease. The PD with an active area of 28$ \times$ 28 mm $^2$ is produced by Hamamatsu company. It is worth noting that the active area of the PD is nearly the same as the readout surface of the smaller scintillator, but is only one half of the bigger one (see Fig.~\ref{setup}). Except the PD, the residual part of the readout surface is packed with the tyvek paper to reflect fluorescent lights back into CsI(Tl). Packing details of CsI(Tl) are shown in Fig.~\ref{CsI} (a).

 \section{$\alpha$ source test}\label{sourcetest}
\begin{figure}
	\centering
		\includegraphics[scale=.28]{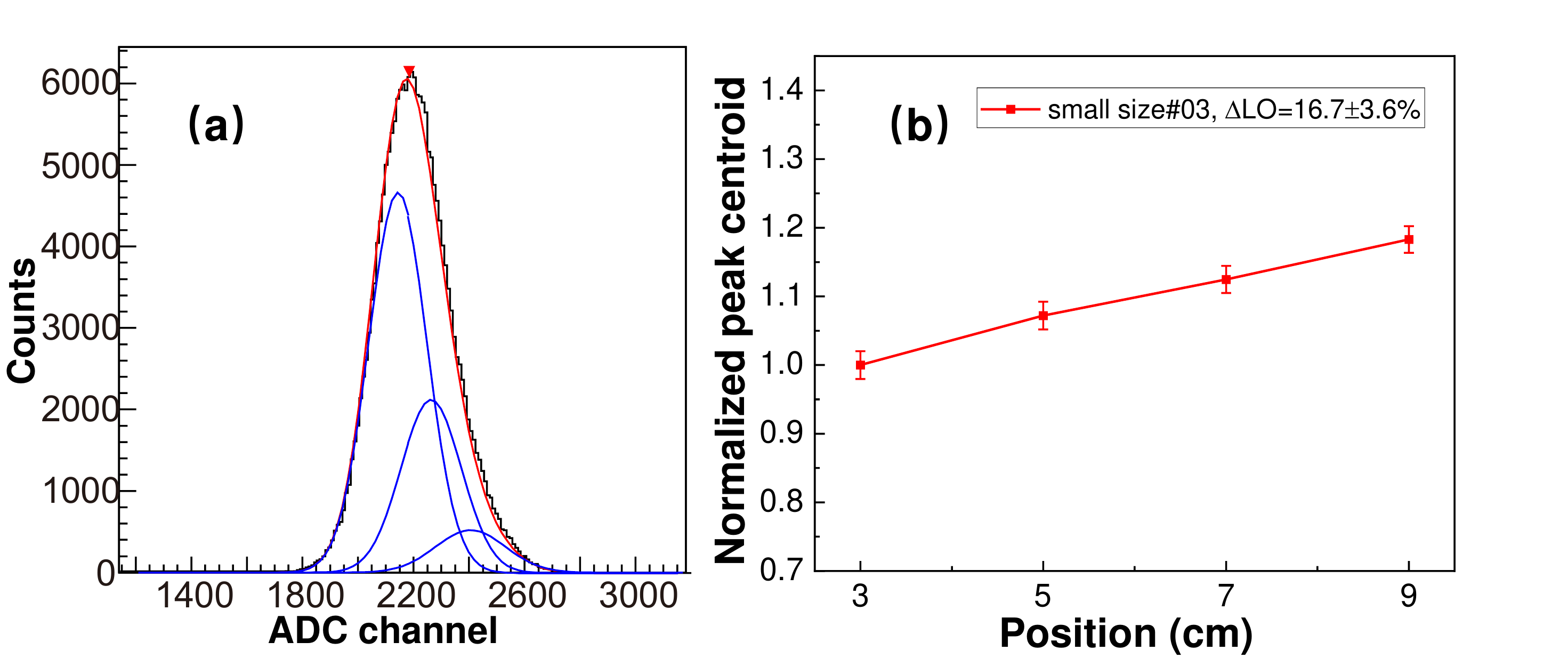}
	\caption{Test results with the three-component $\alpha$ source. (a) The fitted energy spectrum of one CsI(Tl) detector without collimator and (b)one typical light output non-uniformity result with collimator.}
	\label{alphatest}
\end{figure}

The experimental results with a mixing $\alpha$ source which contains three radioactive components, referred to as $^{239}$Pu (5.16 MeV), $^{241}$Am (5.49 MeV), and $^{244}$Ac (5.80 MeV), are shown here. 
Firstly, we placed the source about 3 cm away from the center of CsI(Tl). The incident surface faces to the source directly. The Mylar foil is so thin that $\alpha$ particles can punch through easily. As an example, one typical energy spectrum is shown in Fig.~\ref{alphatest}(a). Due to the poor energy resolution, the energy spectrum is decomposed by three Gaussian functions, where the peak positions are constrained by their known energies and the strength of each peak is 
restrained by the known percent of $\alpha$-particles decaying from each radioactive nucleus. The ADC raw channel of the highest peak centroid is regarded as the pulse height (or electric voltage amplitude), which is proportional to light output. The energy resolution was deduced from the ratio of the full width at half maximum (FWHM) and the centroid of the highest peak. 
The smaller and bigger crystals with the energy resolution under 12\% and 10\%, respectively, are adopted in this step. The poor energy resolution results from the light output non-uniformity of large-size CsI(Tl) and the energy spread of $\alpha$-particles in the Mylar foil with different thickness at different incident angles. Although the resolution is poor, we still found that the energy resolution of the bigger one is obviously better than that of the smaller one, which may be attributed to the better non-uniformity (see test and simulation results below) because the total response of a crystal will be composed of the summed response from all parts.

\begin{figure}
	\centering
		\includegraphics[scale=.26]{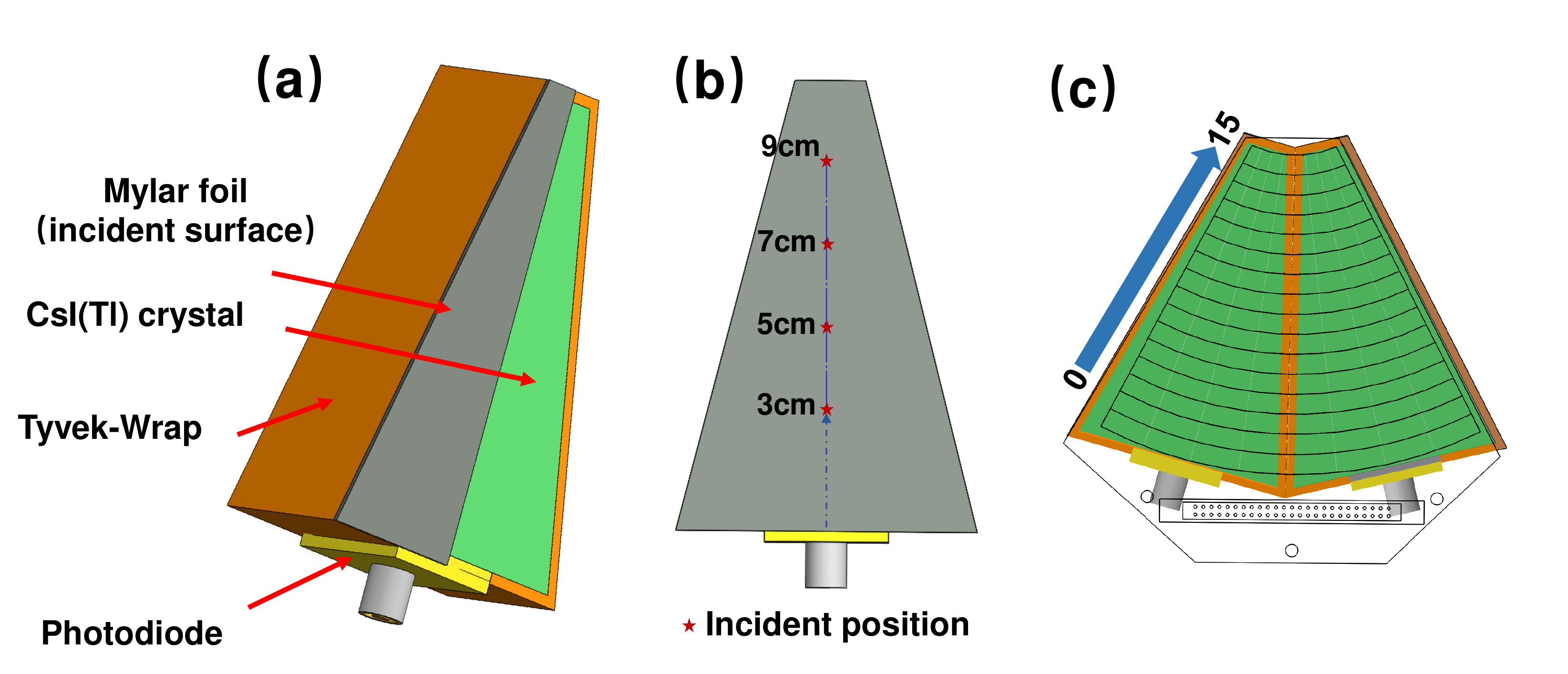}
	\caption{(a)Packing details of the wedge-shaped CsI(Tl). (b) The light outputs are measured at four points along the height of each crystal, locating at 3, 6, 7, 9 cm away from the PD. (c) Configurations for the beam test, in which strip numbers of the ADSSD stand for incident positions of particles.  }
	\label{CsI}
\end{figure}

Secondly, for testing the non-uniformity of light output, we designed a collimator with a 7 mm-radius circular hole and placed it in front of the Mylar foil to limit the incident position of $\alpha$-particles on the CsI(Tl) crystal. As shown in Fig.~\ref{CsI} (b), four different positions locating at 3, 5, 7 and 9 cm away from the PD were tested. One example result is shown in Fig.~\ref{alphatest} (b). It is obvious that the light output increases as the distance becomes farther away from PD or closer to the narrow side. This unexpected trend is consistent with some references \cite{bea1994simulation,knyazev2019properties,AUFFRAY200222}. The light output non-uniformity ($\Delta$LO) in this work is defined as \cite{knyazev2019properties} :

\begin{equation}
      \Delta{ \rm LO} = \frac{C_{\rm max}-C_{\rm min}}{\frac{1}{N}\sum_{i=0}^NC_i}\times100\% 
\end{equation}
where $N$ = 4 is the number of measurement points and $C_i$ is centroid of the highest peak in the energy spectra measured at different incident positions. $C_{\rm max}$ and $C_{\rm min}$ are the maximum and the minimum of $C_i$, respectively.   
The $\Delta$LO value for the smaller and bigger one is 16.7 $\pm$ 3.6 \% and 8.0 $\pm$ 2.8 \%, respectively. It is obvious that the light output non-uniformity of the larger CsI(Tl) is better than the smaller one.

Taking the incident position locating at 5 cm away from the PD for example, the mean energy resolution of the bigger crystal is about 6\%, which is still better than 9\% of the smaller one. The resolution of 6\% is comparable to that reported in Ref.~\cite{knyazev2019properties}. The mean light output of the bigger CsI(Tl) is nearly 18\% lower comparing to the smaller one, which might be attributed to the smaller percent of the PD sensitive area with respect to total area of the readout surface. The light output uncertainty of different crystals at the same incident position (such as the center) is restricted to within 15\% of the mean light output of CsI(Tl)s with the same shape.  

\begin{figure}
	\centering
		\includegraphics[scale=.4]{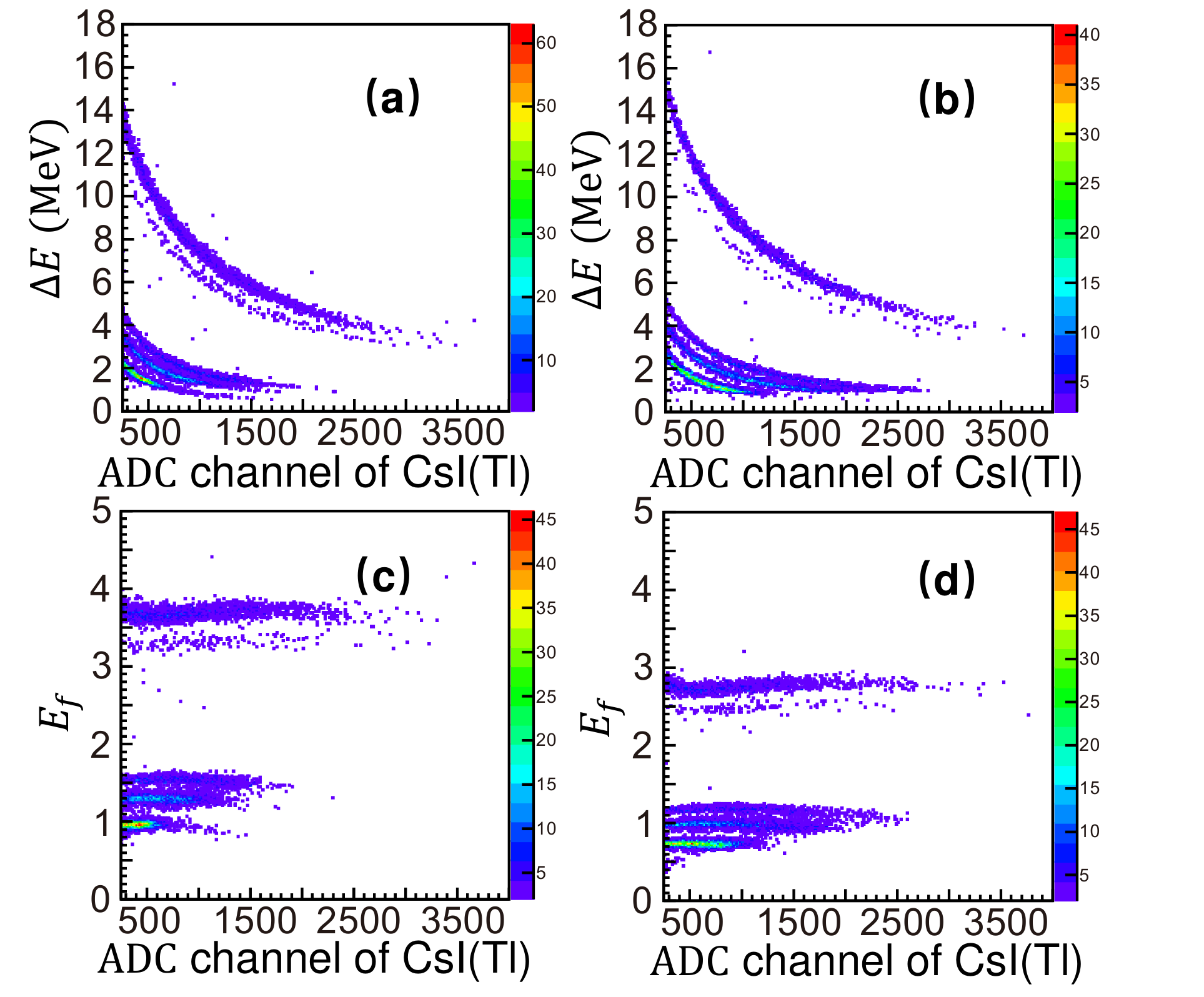}
	\caption{The PID spectra measured by one strip in the ADSSD and (a) one big-size and (b) one small-size CsI(Tl)s. (c) and (d) are the linearized PID, corresponding to (a) and (b), respectively.}
	\label{PID}
\end{figure}
\section{Test with a radioactive beam of $^{15}$C }\label{beamdata}

The beam experiment was performed at Radioactive Ion Beam Line in Lanzhou (RIBLL), Institute of Modem Physics (IMP), China. A $^{15}$C secondary beam at about 27.0 MeV/ nucleon
was produced from a $^{18}$O primary beam at about 60 MeV/nucleon impinging on a $^{9}$Be target with a thickness of 4.5 mm. The secondary beam was purified by a uniform 500-$\micro$m aluminum degrader, and was identified by the TOF provided by two plastic scintillator detectors and energy losses ($\Delta E$) in a large-surface silicon detector (SSD). The average beam intensity and the purity of $^{15}$C were up to 3.4 $\times$ 10$^4$ particles
per second (pps) and 81\%, respectively.

 A 8.63 mg/cm$^2$ (CH$_2$)$_n$ and a 4.95 mg/cm$^2$ (CD$_2$)$_n$ targets were used. Three smaller crystals together with a 148-$\mu$m thick ADSSD are assembled as a $\Delta E$-$E$ telescope (TA1). Another two bigger crystals are assembled as a $E$ detector, which is placed behind a 153-$\mu$m thick ADSSD (TA2). The strip number of the ADSSD relative to the PD is shown in Fig.~\ref{CsI}(c).  
 The telescope TA1 and TA2 were placed downstream about 10.8 cm away from the physical target, covering 20$^\circ$ - 55$^\circ$ in the laboratory frame, to detect the recoil light charged particles, such as $^3$He, $^4$He, $p$, $d$, and $t$.
 
 \subsection{Particle Identification}
 Typical examples of particle identification (PID) spectra with residual energy $E$ measured by one bigger and one smaller CsI(Tl)s are shown in Fig.~\ref{PID}(a) and Fig.~\ref{PID}(b), respectively. Note that $\Delta E$ was detected by one strip in the ADSSD. The strip's position relative to the beam line is the same for these two types of CsI(Tl)s. The ADSSD had been well calibrated strip by strip by using a mixing $\alpha$-source placed in the center of the physical target. The energy resolution for one strip is about 1.8\% for the 5.486-MeV $\alpha$ particle.
  It is obvious that the $Z$ = 1 and $Z$ = 2 isotopes are identified clearly by both telescopes , TA1 and TA2, using the $\Delta E-E$ technique. In order to evaluate and compare the PID capability of different CsI(Tl) crystals, a linearization method was introduced~\cite{uroic2015improvements}: 

\begin{equation}
  E_f =\sqrt{\Delta E\cdot(a_0\cdot{ch}+a_1)+a_2\Delta E^2}+a_3(a_0\cdot{ch}+a_1)
\end{equation}
 where $E_f$ is a constant while using proper parameters of $a_0$, $a_1$, $a_2$, $a_3$, which are obtained from fitting the isotopic bands in Fig.~\ref{PID}. The corresponding linearized results of Fig.~\ref{PID}(a) and Fig.~\ref{PID}(b) are shown in Fig.~\ref{PID}(c) and Fig.~\ref{PID}(d), respectively.
 
 \begin{figure}
	\centering
		\includegraphics[scale=.6]{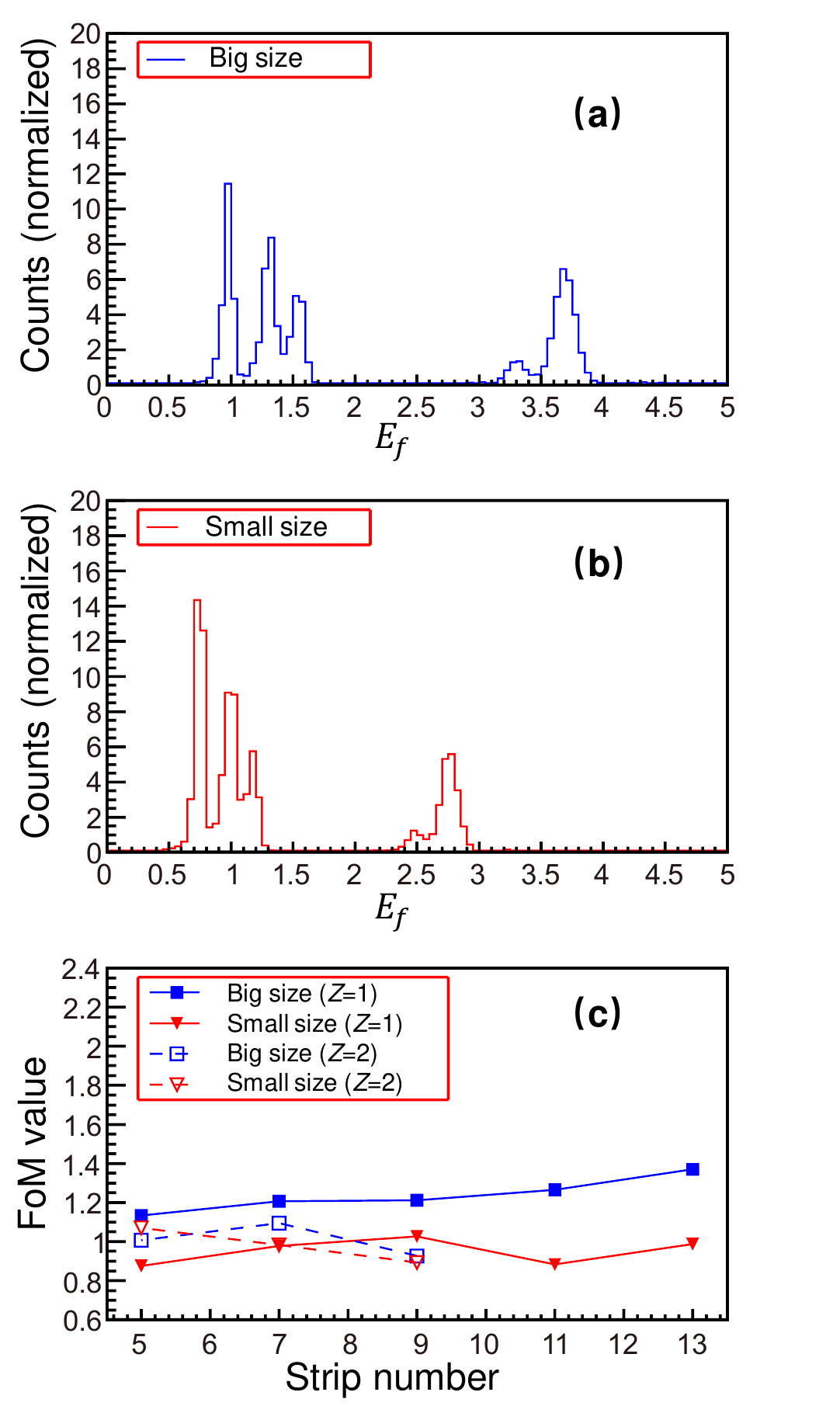}
	\caption{The projection of PID spectra measured by one big-size (a) and one small-size (b) crystals . (C) The FoM values as a function of strip numbers of the corresponding ADSSD.}
	\label{projection}
\end{figure}

The projection of PID spectra through y-axis for different CsI(Tl)s are shown in Fig.~\ref{projection}. The isotopic discrimination capability can be quantitatively evaluated by the average value of "Figure of Merit" (FoM), defined as~\cite{winyard1971pulse}:

\begin{equation}
\rm FoM=\frac{\lvert \overline{PID_2}-\overline{PID_1}\rvert}{FWHM_1+FWHM_2}
\end{equation}
where $\rm \overline{PID_1}$ and $\rm\overline{PID_2}$ are the centroids of two neighboring isotope peaks, while $\rm FWHM_1 $ and $\rm FWHM_2$ are their FWHMs. It is said well-separated if FoM is greater than 0.7~\cite{CARBONI2012251}. The mean FoM values for the $Z$ = 1 and $Z$ = 2 isotopes as a function of the strip number in the ADSSD are shown in Fig.~\ref{projection}(c). All the FoM values are greater than 0.9. Thus, it is good enough for all the crystals to identify the light charged particles recoil from various direct reactions, such as $^3$He, $^4$He, $p$, $d$, and $t$ from ($d$, $^3$He), ($d$, $^4$He), ($p$, $p$'), ($d$, $d$'), and ($d$, $t$), respectively. For the hydrogen isotopes, it was found that the FoM value of the bigger crystal is higher than that of the smaller one, indicating that the big-size CsI(Tl) has a better capability to discriminate hydrogen isotopes than the small-size one. This conclusion is independent of the strip number, which stands for the incident position (seeing Fig.~\ref{CsI} (c)). For the helium isotopes, the limited experimental data show that it is not in conflict with this conclusion. 

\subsection{Energy Calibration of CsI(Tl)}
We used the beam test data to check the light output non-uniformity in more detail. First of all, we should calibrate the CsI(Tl) crystal in order to conveniently choose the same particle with a certain energy. It is impossible to calibrate CsI(Tl) using a low-energy $\alpha$-source due to the non-linearity and particle-dependence of energy response. 
As shown in Fig.~\ref{PID}, the $\Delta E$ detector, ADSSD, were well calibrated by using a mixing $\alpha$ source. Using the $\Delta E$ value together with the known thickness of the silicon detector, the total kinetic energy of the isotope impinging on the ADSSD layer ($E_{\rm {total}}$) can be deduced by a numerical inversion of Ziegler's energy loss tables, as described in Ref.~\cite{DELLAQUILA2019162}.
The residual energy $E$ is easily calculated by the relationship of $E = E_{\rm{total}}-\Delta E$. The different thickness of each strip at distinct incoming angles with respect to the beam direction was taken into consideration during the calibration procedure of CsI(Tl). Comparing to this kind of thickness variation, the thickness non-uniformity of one strip in the ADSSD is smaller and is ignored in this paper. 

\begin{figure}
	\centering
		\includegraphics[scale=.34]{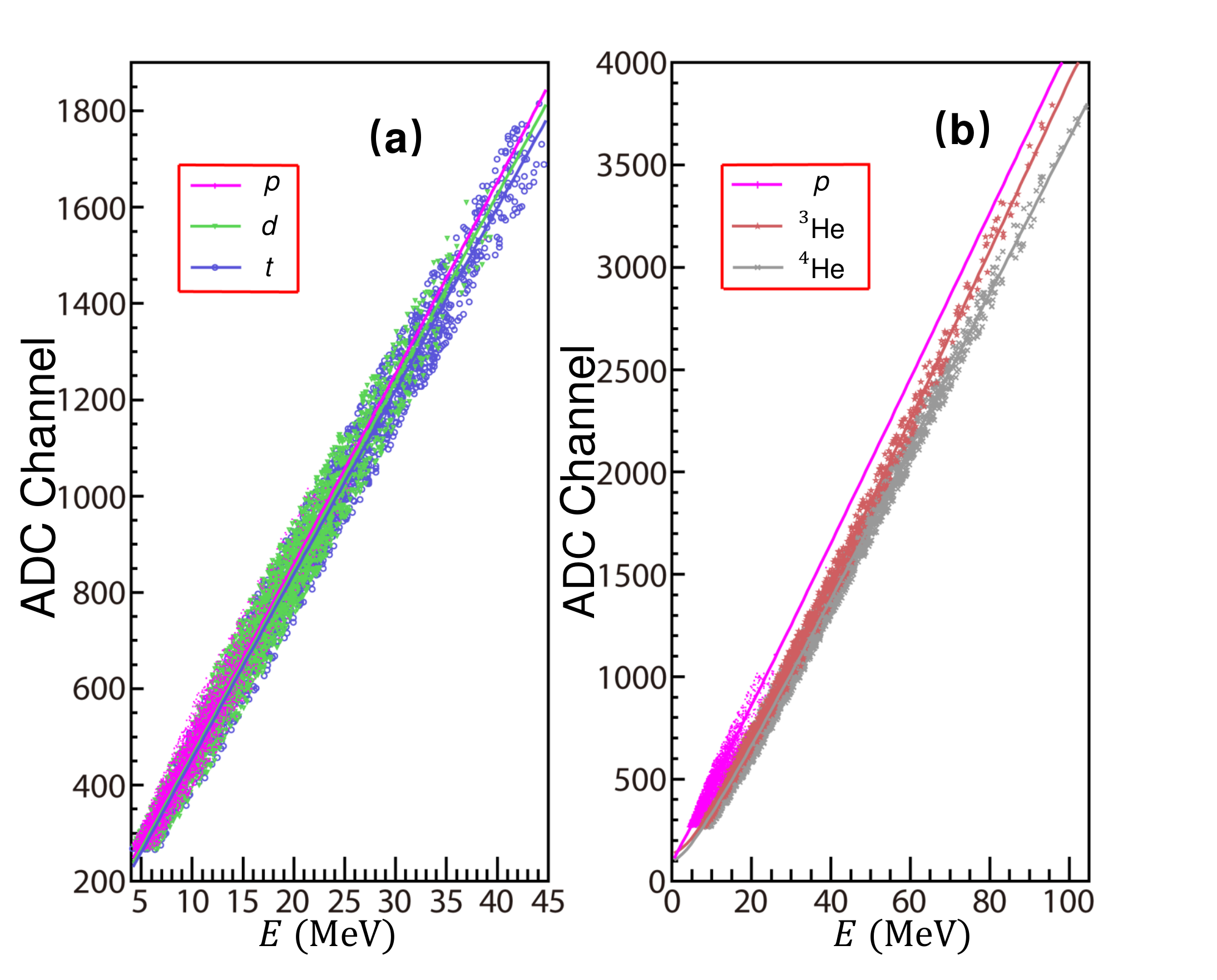}
	\caption{The light output as a function of incident energy for $Z$ = 1 (a) and $Z$ = 2 (b) isotopes identified by one big-size CsI(Tl) and the 9th strip in its corresponding ADSSD. }
	\label{calibration}
\end{figure}

The calibration formula applied for the $Z = 1$ isotopes is described by the following empirical parametrization ~\cite{DELLAQUILA2019162},

\begin{equation}
    L(E, Z = 1, A) = a_0E^{a_1}+a_2
\end{equation}
where $a_0$ is a gain factor, $a_1$ is an empirical non-linearity parameter, and $a_2$ is an adding parameter to fit the experimental data better. The parameters were obtained from fitting each isotopic band individually. The typical light output as a function of incident energy is given in Fig.~\ref{calibration}(a). In a wide energy range (8-80 MeV), the light output does not exhibit a significant non-linearity and a obvious separation among protons, deuterons and tritons. This trend is consistent with that shown in Refs.~\cite{DELLAQUILA2019162,AIELLO199650}.

For the heavier helium isotopes, another formula is adopted ~\cite{DELLAQUILA2019162}: 
\begin{equation}
    L(E, Z = 2, A) = a_0 ( E - a_1AZ^2log(\frac{E+a_1AZ^2}{a_1AZ^2}))+a_2
\end{equation}
where $a_0$, $a_1$ and $a_2$ are parameters obtained from the fit of experimental data of individual helium isotopes. $a_0$ is related to the scintillation efficiency of the crystal while $a_1$ is a non-linear quenching factor dominating the lower energy region. $a_2$ is an adding parameter in order to fit the experimental data better. Comparing to protons, the fitted results of helium isotopes are shown in Fig.~\ref{calibration}(b). We found that the light outputs for helium isotopes are consistently lower and the non-linearity is more apparent with respect to the one measured for protons at lower energies. The light output for $^3$He is obviously larger relative to $^4$He, especially at incident energy higher than 60 MeV. 
These trends are similar to the ones described in Ref.~\cite{DELLAQUILA2019162} and can be interpreted by the Birks' function of inorganic scintillation \cite{Briks}

\begin{equation}
    dL/dE =\frac{S}{1+kB|dE/dx|}
\end{equation}
where $S$, $kB$ and $dE/dx \propto AZ^2/E$ are the scintillation efficiency, quenching factor and stopping power, respectively. The scintillation efficiency, $S$, is almost a constant at sufficiently high energies, but is reduced as the energy decreases down to less than 20 MeV, leading to the non-linearity at low energies. In the case of small $S$ and small $dE/dx$ for the $Z$ = 1 isotopes , the light output is nearly linear, see Fig.~\ref{calibration} (a). For the $Z$ = 2 isotopes, the part of $kB \times dE/dx$ will play a more important role at lower energies than that for the hydrogen isotopes ($Z$ = 1) because the $dE/dx \propto AZ^2/E $ value is larger. As a result, 
the non-linearity for the helium isotopes ($Z$ = 2) at energy lower than 20 MeV is more apparent (Fig.~\ref{calibration} (b)). The stopping power, $dE/dx \propto AZ^2/E$, of $^3$He is smaller than that of $^4$He, resulting in a larger light output for $^3$He at all incident energies. 

\subsection{Non-uniformity of light output}
\begin{figure}
	\centering
		\includegraphics[scale=.55]{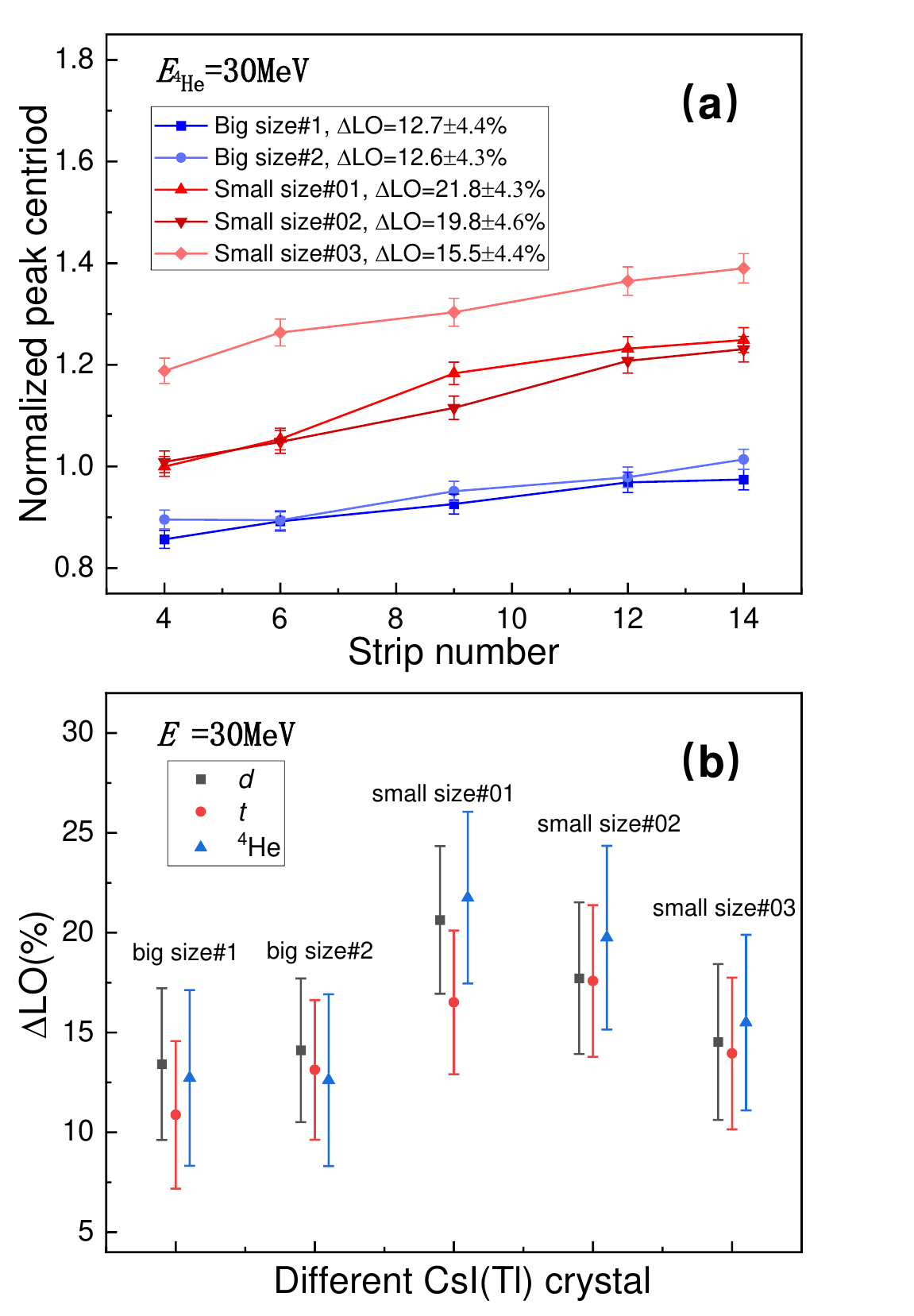}
	\caption{(a) Light outputs of five CsI(Tl) crystals change as a function of strip number in the ADSSD, which represents the hit position. (b) $\Delta$LO values measured with different particles for five CsI(Tl) detectors .}
	\label{nonuniformity}
\end{figure}

After calibration, we can investigate the light output non-uniformity of the wedge-shaped CsI(Tl) by using a certain particle with definite incident energy. Here, as shown in Fig.~\ref{CsI}(c), the incident position was tagged by the strip number in the ADSSD. 
First of all, tritons with the energy of 30.0 $\pm$ 0.5 MeV deposited in CsI(Tl) were chosen to compare the non-uniformity of different crystals. The detailed results are shown in Fig.~\ref{nonuniformity}(a). It was found that the light output changes with the hit position, which is consistent with the test result with $\alpha$ source, see Fig.~\ref{alphatest}(b). Three smaller crystals have a general worse non-uniformity than two bigger crystals have while two bigger crystals have a worse light output.  In addition, the $\alpha$-particles with even lower energies of 10.0 $\pm$ 0.2 MeV were also opted to check the hit position dependence of light output. The results are nearly the same for the same CsI(Tl), which demonstrates that this light output non-uniformity of CsI(Tl) seems to be particle-independent.  
Furthermore, we extracted the $\Delta$LO values for each crystal with deuterons, tritons, and alpha-particles at the same incident energy of 30 MeV. The results are shown in Fig.~\ref{nonuniformity}(b). 

 Generally, the $\Delta$LO values for three small-size crystals are obviously larger than those for two big-size CsI(Tl)s. For the same crystal, the $\Delta$LO value is nearly independent of different incident particles.
 It indicates that light propagation in different wedge-shaped CsI(Tl)s and light collection by the PD rather than the incident particles, are the main factors that affect light output non-uniformity.

\section{Geant4 Simulation}
 Light propagation and light collection in these two different types of CsI(Tl)s were simulated by using the Geant4 package (version of 10.1.1)~\cite{geant4}. The photon yield (scintillation efficiency) is set to be 65000/MeV, and the light absorption length through the propagation path is 35 cm~\cite{knyazev2019properties}. We assumed that protons with an energy of 30 $\pm$ 0.5 MeV produce fluorescence photons over different positions along the height of CsI(Tl), see Fig.~\ref{CsI}(b). Detailed size and packing materials of CsI(Tl)s are the same as real ones. The intrinsic property of CsI(Tl)s with two different shapes are set to the same to each other.

\begin{figure}
	\centering
		\includegraphics[scale=.32]{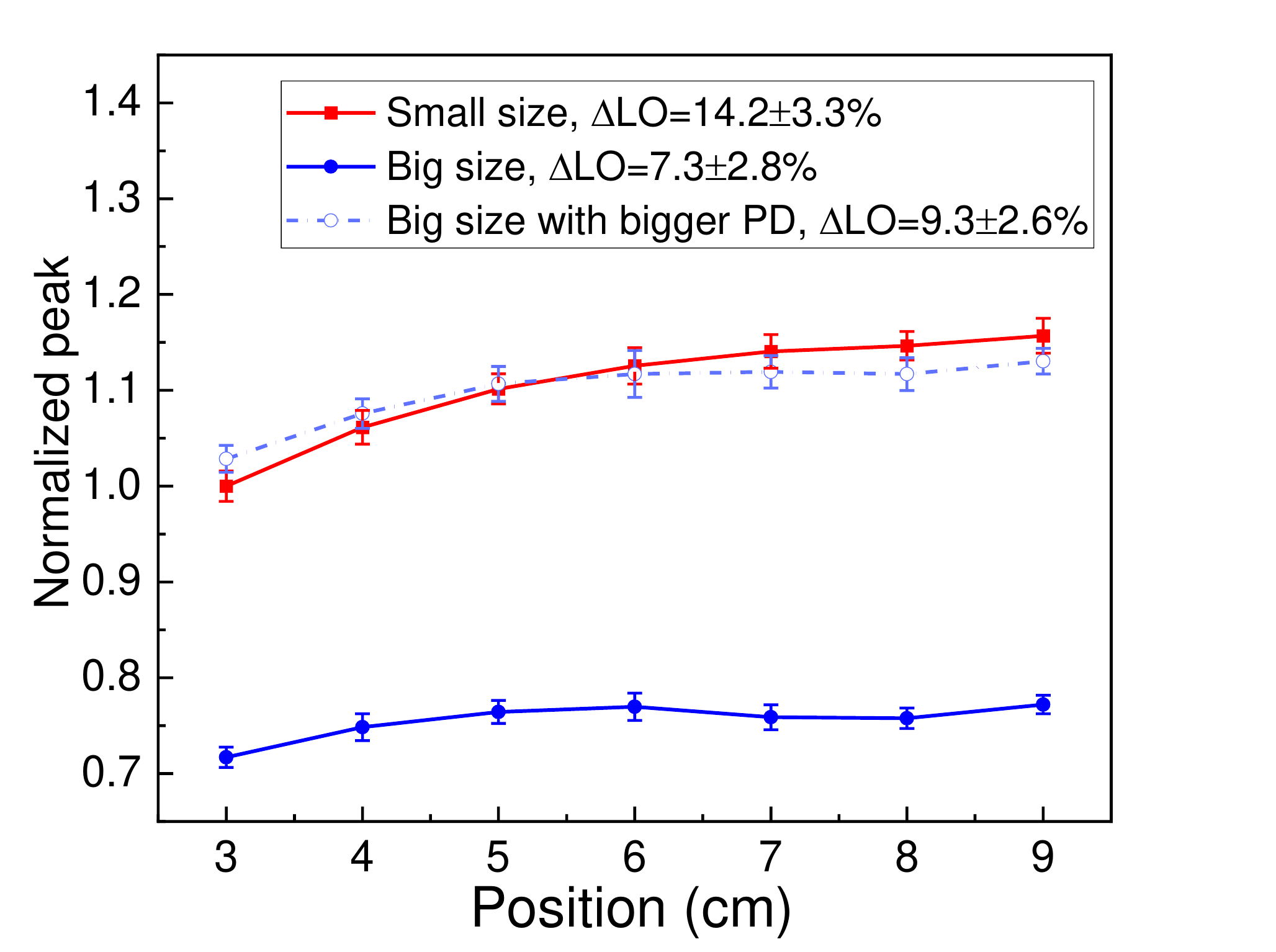}
	\caption{Simulation results of light output variation over different hit positions. }
	\label{simulation}
\end{figure}

The simulated results are shown in Fig.~\ref{simulation}, the trend of light output is consistent with the $\alpha$- source and beam test results. For the big-size CsI(Tl), the ascending slope of light output  (blue solid circles) is smaller with respect to the small-size one (red solid squares), which results in a lower $\Delta$LO value of 7.0\% . However, the absolute amount of light collected by the PD coupled to the bigger CsI(Tl) is obviously lower than the smaller crystal. If we increase the sensitive area of the PD up to the same percent of smaller CsI(Tl), the light collection amount (blue empty triangles) increases, but the $\Delta$LO value only changes a little, from 7.0\% to 9.2\%. Therefore, in the future, if possible, it is better to adopt the large-area PD, which can improve light collection amount but deteriorate the non-uniformity a little. 

\begin{figure}
	\centering
		\includegraphics[scale=.2]{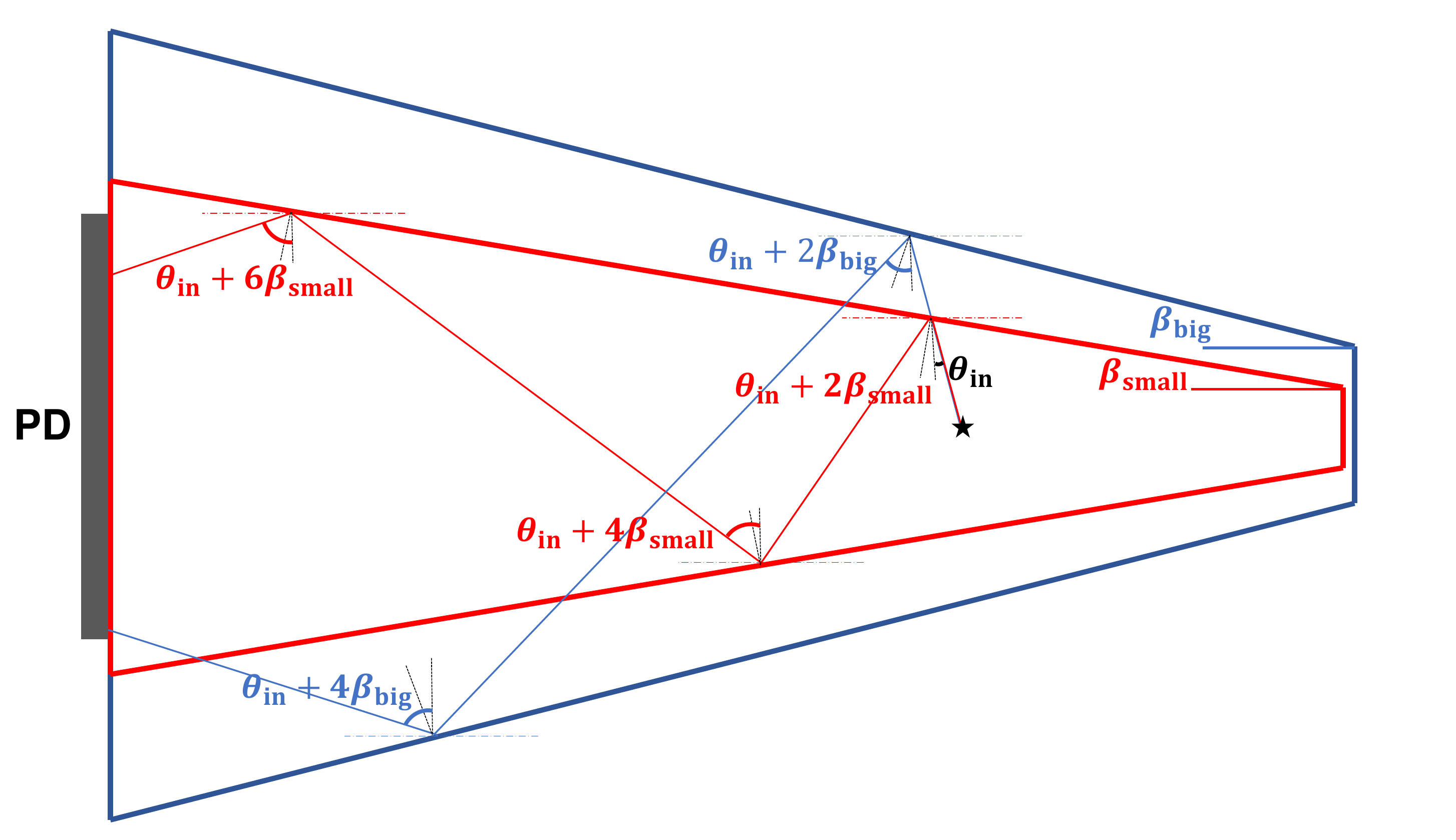}
	\caption{Different reflections of fluorescence light on the side-surface of big-size (blue) and small-size (red) CsI(Tl)s, where $\beta$ and $\theta_{\rm in}$ represent the tapering angle of crystal and the incident angle of fluorescence light with respect to the horizontal (dashed) line, respectively. }
	\label{lightreflection}
\end{figure}

The light output non-uniformity is dependent on the intrinsic property, the surface roughness, as well as the geometry shape of the CsI(Tl) crystals \cite{bea1994simulation,knyazev2019properties,AUFFRAY200222}. In addition, it is also affected by the light attenuation through the propagation path \cite{knyazev2019properties}, and by the sensitive area of the PD which collects the light signal. The tapered crystal, such as wedge-shaped in the present paper, can focus light due to the specular light reflection on the surface of a tapered crystal \cite{bea1994simulation,knyazev2019properties,AUFFRAY200222}. As shown in Fig.~\ref{lightreflection}, 
for each reflection, the light with an incident angle of $\theta_{\rm in}$ relative to the horizontal (dashed) line, will reflect into an angle of $\theta$ + $2\beta$, where $\beta$ is the tapering angle of CsI(Tl)\cite{knyazev2019properties}. This increase of outgoing angle will allow the PD to collect light emitting to larger solid angles. This increase is related to the product of tapering angle $\beta$ and reflection number $n$. Thus, in the case of different positions over one wedge-shaped crystal, more light will transmit to the readout surface as the distance between the incident position and the readout surface increases because the reflection number $n$ ascends.
Comparing the bigger CsI(Tl) with the smaller one, on one side, the larger $\beta$ angle strengthens the focusing effect, but on the other side, the less reflection number $n$ on the surface weakens this focusing effect. At the same time, the length of propagation path also affects the light output amount. As a comprehensive result of various factors mentioned above, for the large-size CsI(Tl), the light output non-uniformity is better, leading to that the energy resolution and the PID capability are better than those of the small-size ones. Therefore, the large-size CsI(Tl) with the bigger PD is more appropriate to form the telescope TA. The simulation results also show that the tapering shape rather than different intrinsic property of CsI(Tl) along its height (such as non-uniformity of the doped Tl) is the main reason for the light output non-uniformity.  

\section{Summary}

The properties of wedge-shaped CsI(Tl)s with two different shapes are reported in this paper. They are installed behind the ADSSD to form the $\Delta E - E$ telescope, allowing to identify the light charged particles produced in direct nuclear reaction experiments. Their energy resolution, particle identification capability, and light output non-uniformity were measured by using a mixing $\alpha$-source and a radioactive beam of $^{15}$C at an incident energy of 27.0 MeV/nucleon. We found advantages of applying the big-size crystals to form the ADSSD-CsI(Tl) telescope, such as better energy resolution, smaller light output non-uniformity, and better particle identification capability. The light output non-uniformity of these wedge-shaped CsI(Tl)s can be explained based on the Geant4 simulation results. Furthermore, the simulation results are in favor of using the PD with bigger sensitive area, which allows to increase the light collection amount but without deteriorating the light output non-uniformity.

\section*{\label{sec:level1}Acknowledgments}
We gratefully thank the accelerator group of institute modern physics (IMP) for providing $^{18}$O primary beam and the RIBLL collaboration for supplying a lot of electronics. This work was supported by
the National Natural Science Foundation of China (Contract Nos.U1867214,11775004,11875074,  11961141003).

\appendix


\printcredits

\bibliographystyle{elsarticle-num}
\bibliography{cas-refs}


\bio{}
\endbio


\endbio

\end{document}